\documentclass[twocolumn,showpacs,preprintnumbers,amsmath,amssymb]{revtex4}

\usepackage{graphicx}
\usepackage{dcolumn}
\usepackage{bm}

\begin{document}


\title{%
Evidence for $s$-wave superconductivity 
in the new $\beta$-pyrochlore oxide RbOs$_2$O$_6$ 
}%

\author{K. Magishi$^{1,2}$, J.L. Gavilano$^1$, B. Pedrini$^1$, 
J. Hinderer$^1$, M. Weller$^1$, H.R. Ott$^1$, 
S.M. Kazakov$^1$, and J. Karpinski$^1$ } 
%
\affiliation{$^1$Laboratorium f\"{u}r Festk\"{o}rperphysik, 
ETH-H\"{o}nggerberg, CH-8093 Z\"{u}rich, Switzerland}
\affiliation{$^2$Faculty of Integrated Arts and Sciences, 
The University of Tokushima, Tokushima 770-8502, Japan}

\date{\today}

\begin{abstract}
We report the results of $^{87}$Rb NMR measurements on RbOs$_2$O$_6$, 
a new member of the family of the superconducting pyrochlore-type oxides 
with a critical temperature $T_c$ = 6.4 K. 
In the normal state, 
the nuclear spin-lattice relaxation time $T_1$ obeys 
the Korringa-type relation $T_1T$ = constant 
and the Knight shift is independent of temperature, 
indicating the absence of strong magnetic correlations. 
In the superconducting state, 
$T_1^{-1}(T)$ exhibits a tiny coherence enhancement just below $T_c$, 
and decreases exponentially with further decreasing temperatures. 
The value of the corresponding energy gap is 
close to that predicted by the conventional weak-coupling BCS theory. 
Our results indicate that 
RbOs$_2$O$_6$ is a conventional $s$-wave-type superconductor. 
\end{abstract}

\maketitle


Geometrical frustration of spin systems has recently attracted much attention 
because, instead of long-range magnetic order, 
novel ground states, including unconventional superconductivity, 
may be adopted\cite{Pyro-1,Pyro-2}. 
Pyrochlore-type oxides with tetrahedral networks of magnetic ions, 
the so-called pyrochlore lattice, are well known physical realizations of 
geometrically frustrated magnetic systems. 
This class of materials has recently received enhanced attention 
because superconductivity was found in Cd$_2$Re$_2$O$_7$ 
below $T_c$ = 1 K\cite{Cd-1,Cd-2,Cd-3}. 
The mechanism causing this superconductivity appears to be conventional 
and the physical properties are compatible with expectations based on 
the weak-coupling BCS theory\cite{Cd-4}. 
It was suggested that the onset of superconductivity 
might be related to an unusual structural phase transition\cite{Cd-5,Cd-6}. 
Another superconducting pyrochlore, KOs$_2$O$_6$, 
seems to be a quite different case. 
This material crystallizes in the $\beta$-pyrochlore structure 
and exhibits superconductivity below $T_c$ = 9.6 K\cite{K-1}. 
In contrast to Cd$_2$Re$_2$O$_7$, 
KOs$_2$O$_6$ does not exhibit any structural phase transition near $T_c$. 
This suggests that the geometrical frustration 
persists to temperatures below $T_c$. 
The upper critical magnetic field $H_{c2}$ appears to exceed 
the Pauli limit expected for conventional superconductivity, 
This observation was interpreted as a sign for 
unconventional superconductivity in KOs$_2$O$_6$\cite{K-2}. 
The results of recent $\mu$SR experiments were interpreted 
as to strongly suggest that the superconducting state of KOs$_2$O$_6$ 
is unconventional, characterized by gap nodes\cite{K-3}. 
Analogous experiments invoking Cd$_2$Re$_2$O$_7$\cite{Cd-7} 
revealed an isotropic gap for that material, instead. 

Very recently, a new ternary compound RbOs$_2$O$_6$ 
with the $\beta$-pyrochlore structure, the same as KOs$_2$O$_6$, 
has been discovered\cite{Rb-1}. 
This compound exhibits superconductivity below $T_c$ = 6.4 K\cite{Rb-1,Rb-2}. 
The results of recent specific-heat\cite{Rb-2} 
and magnetic-field penetration depth measurements\cite{Rb-3} 
were claimed to indicate a conventional BCS-type behavior 
of this superconducting state. 
The zero-temperature upper critical magnetic field $H_{c2}$ ($\sim$ 6 T), 
extracted from specific-heat measurements on RbOs$_2$O$_6$, 
is lower than the Pauli-limiting field\cite{Rb-2}, 
in contrast to the above cited claims for KOs$_2$O$_6$. 
From the results of electrical resistivity measurements, however, 
the value of $H_{c2}$ for RbOs$_2$O$_6$ at zero temperature 
was claimed to be larger than the Pauli limit of 12 T\cite{Rb-1}. 
This estimated value for the Pauli limit may, however, be substantially 
modified towards higher values by spin-orbit interactions, 
as suggested by the results of band-structure calculations 
for the related pyrochlore-type oxides 
Cd$_2$Re$_2$O$_7$ and Cd$_2$Os$_2$O$_7$\cite{Harima}. 

In this report, we present the results of magnetic susceptibility 
and $^{87}$Rb NMR measurements on samples of polycrystalline RbOs$_{2}$O$_{6}$ 
in both the normal and the superconducting state. 
Our results support the view of conventional superconductivity 
in this material. 


The samples were prepared from polycrystalline material of RbOs$_{2}$O$_{6}$, 
synthesized from the starting materials OsO$_{2}$ and Rb$_2$O. 
The experimental details of the synthesis and the purification 
of RbOs$_{2}$O$_{6}$ are described elsewhere\cite{Rb-4}. 
The material was confirmed 
to have the correct structure by X-ray diffraction. 
Most of the reflections could be indexed 
on the basis of a pyrochlore unit cell 
with a lattice parameter $a$ = 10.1137(1) $\AA$. 
A small amount of OsO$_2$ (less than 5 $\%$) was detected as an impurity. 
Our NMR experiments probe the local environment of the Rb ions. 
The Rb cations occupy the $8b$ site in the pyrochlore lattice, 
which provides a local environment with cubic symmetry. 

The magnetic susceptibility $\chi(T) = \frac{M(T)}{H}$, 
where $M(T)$ represents the temperature dependent magnetization, 
was measured upon cooling the sample at temperatures between 300 K and 2 K 
in external magnetic fields $\mu_0H$ of 50 G and 2.94 T, 
using a SQUID magnetometer. 
The NMR measurements were performed at temperatures between 0.4 K and 35 K 
in an external magnetic field of 2.9427 T 
using a standard phase-coherent-type pulsed spectrometer. 
The $^{87}$Rb NMR spectra were obtained by Fast-Fourier-Transfomation (FFT) 
of the spin-echo signals, following a $\frac{\pi}{2} - \pi$, 
$rf$ pulse sequence. 
The nuclear spin-lattice relaxation time $T_{1}$ was measured 
by the saturation recovery method, 
where the spin-echo signals were measured 
after the application of a comb of $rf$ pulses and subsequent variable delays. 


%
\begin{figure}
\includegraphics[width=0.8\linewidth]{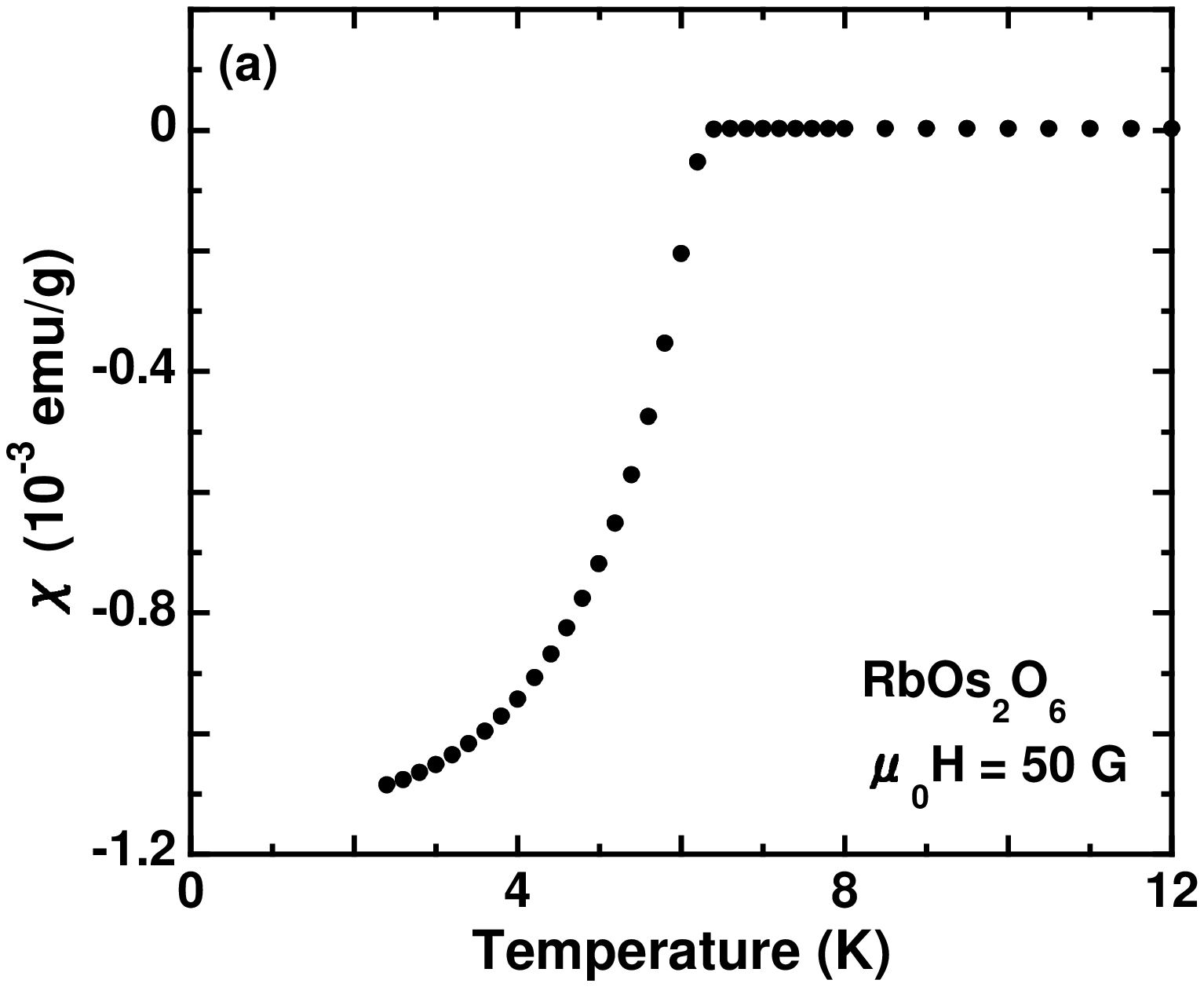}
\end{figure}
\begin{figure}
\includegraphics[width=0.8\linewidth]{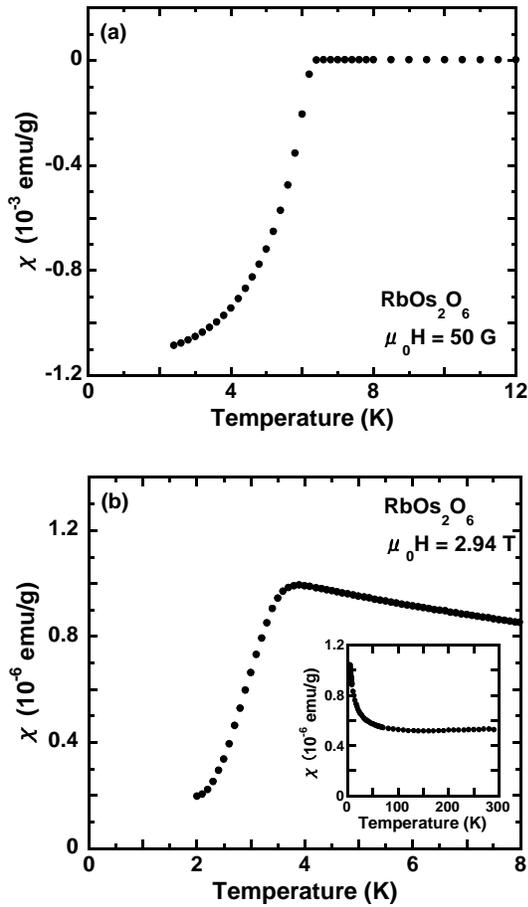}
\caption{\label{fig:epsart} 
Temperature dependences of the magnetic susceptibility $\chi(T)$ 
measured on a powdered sample of RbOs$_2$O$_6$ 
in external magnetic fields of (a) 50 G and (b) 2.94 T. 
Note the very different vertical scales. 
The sample was cooled in the field. 
Inset : Temperature dependence of $\chi(T)$ above $T_c$ 
in an external magnetic field of 2.94 T. }
\end{figure}

Figures 1(a) and 1(b) show the temperature dependences of 
the magnetic susceptibility $\chi(T)$ of a powdered sample of RbOs$_2$O$_6$. 
The superconducting transition is reflected in the onset of 
a large diamagnetic signal due to the Meissner effect (Fig. 1(a)). 
In case of $\mu_0H$ = 50 G, 
$\chi(T)$ reflects the onset of diamagnetism at 6.4 K. 
As usual, increasing external magnetic fields shift the transition 
to lower temperatures\cite{Rb-1,Rb-2}. 
In an external magnetic field $\mu_0H$ = 2.94 T, 
the same that we used in our NMR measurements, 
the $\chi(T)$ data reveal the onset of superconductivity 
at $T_c$ = 3.8 K (Fig. 1(b)). 
This value is consistent with 
the results of previous specific-heat measurements\cite{Rb-2}. 
The inset of Fig. 1(b) shows the temperature dependence of $\chi(T)$ 
of RbOs$_2$O$_6$ in the normal state for $\mu_0H$ = 2.94 T. 
At temperatures exceeding 100 K, 
$\chi(T)$ is, to a good approximation, temperature-independent. 
Upon cooling to below 50 K, 
the susceptibility increases gradually, 
such that $\chi(T) = \chi_0 + \frac{C}{T}$, where $C$ is the Curie constant 
and $\chi_0$ = 4.8 $\times 10^{-7}$ emu/g 
is the temperature-independent susceptibility. 
For common metals, $\chi_0 = \chi_{Pauli} + \chi_{Landau} + \chi_{shell}$, 
where for free electrons $\chi_{Landau} = -\frac{1}{3}\chi_{Pauli}$. 
Assuming that the core-electron diamagnetism term $\chi_{shell}$ 
is negligibly small, $\chi_{Pauli} = \frac{3}{2}\chi_0$ and 
we can calculate the electronic density of states at the Fermi surface 
from $\chi_{Pauli} = \mu_{\rm B}^2D(E_F)$. 
Using this relation, we obtain $D(E_F)$ = 1.38 states/eV$\cdot$atom. 
This value of $D(E_F)$ implies that the electronic specific-heat coefficient 
$\gamma = \frac{\pi^2k_{\rm B}^2D(E_F)}{3}$ = 31 mJ$\cdot$mol$^{-1}$K$^{-2}$, 
in very good agreement with the value 
estimated from the result of specific-heat measurements\cite{Rb-2}. 
The Curie-type upturn of $\chi(T)$ at low temperatures 
is attributed to the presence of a small concentration of impurity moments. 
The effective paramagnetic moment deduced from the Curie constant 
is small, of the order of 0.1 $\mu_{\rm B}$/Os. 
Thus, the intrinsic behavior of $\chi(T)$ of RbOs$_2$O$_6$ in the normal state 
is that of a simple metal, consistent with the results of 
measurements of the Knight shift and the nuclear spin-lattice relaxation rate, 
to be discussed below. 


%
\begin{figure}
\includegraphics[width=0.8\linewidth]{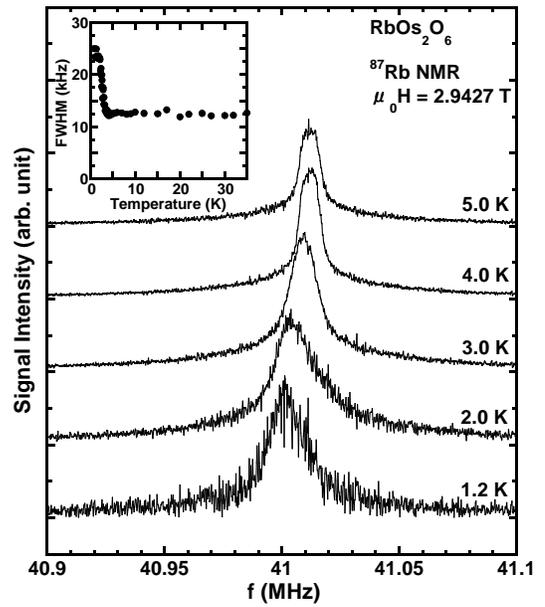}
\caption{\label{fig:epsart} 
Evolution of the FFT spectra of $^{87}$Rb NMR 
in an external magnetic field of 2.9427 T between 1.2 K and 5.0 K. 
Inset : Temperature dependence of the full-width at half-maximum (FWHM) 
of the signal. 
}
\end{figure}

Figure 2 shows the evolution of the FFT spectra of $^{87}$Rb NMR 
at low temperatures. 
The spectrum contains a single resonance line. As described above, 
the Rb nuclei occupy highly symmetrical sites and thus, 
the influence of the quadrupole interaction is quenched 
($I=\frac{3}{2}$ for the $^{87}$Rb nucleus). 
The inset of Fig. 2 shows the temperature dependence of 
the full-width at half-maximum (FWHM) of the $^{87}$Rb NMR spectrum. 
In the normal state, this width is of the order of 12 kHz and 
independent of temperature. 
In the superconducting state, 
the NMR spectrum broadens appreciably 
due to a distribution of local fields produced by the vortex lattice. 
The inhomogeneous broadening of the NMR line of type-II superconductors 
can approximately be calculated as 
$\Gamma \sim \frac{\phi_0}{\lambda^2(16\pi^3)^{1/2}}$, 
the square root of the second moment of the expected field distribution 
due to the vortices\cite{Pincus}. 
With $\lambda = 4100 \AA$ 
as the zero-temperature London penetration depth\cite{Rb-3} 
and $\phi_0 = \frac{hc}{2e}$ as the flux quantum, 
we calculate $\Gamma \sim$ 5.5 Oe $\sim$ 8 kHz at $T$ = 0 K. 
This value is close to the observed total enhancement of the line width of 
approximately 12 kHz. 


%
\begin{figure}
\includegraphics[width=0.8\linewidth]{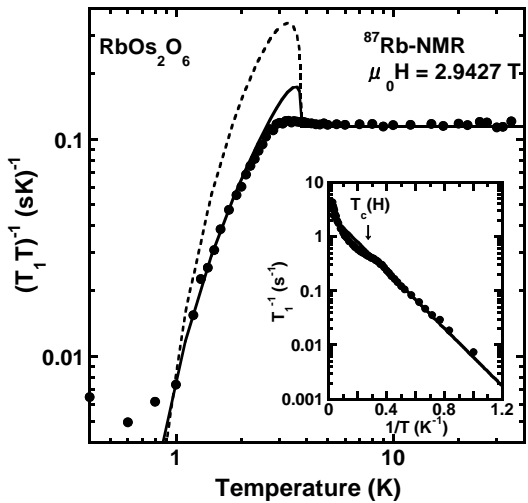}
\caption{\label{fig:epsart} 
Temperature dependence of $(T_1T)^{-1}$. 
Dotted and solid lines represent the BCS expectation for an isotropic 
and anisotropic gap, respectively (see text). 
Inset : Semilogarithmic plot of $T_1^{-1}$ vs $\frac{1}{T}$. 
The solid line represents the relation 
$T_1^{-1} \propto \exp(-\frac{\Delta(0)}{k_{\rm B}T})$ 
with $\frac{\Delta(0)}{k_{\rm B}}$ = 6.1 K. 
}
\end{figure}

Next, we consider the nuclear spin-lattice relaxation. 
Figure 3 shows the temperature dependence of $(T_1T)^{-1}$. 
The $T_1$ measurements were made at the peak positions 
of the resonance signals, but 
the employed $rf$ pulses were short enough to irradiate the entire NMR line. 
The observed magnetization recovery (data not shown) 
followed a single-exponential curve. 

In the normal state, $T_1^{-1}(T)$ 
obeys the Korringa relation $(T_1T)_n^{-1}$ = 0.117 (sK)$^{-1}$, 
as expected for simple metals 
and indicating the absence of significant magnetic interactions 
in RbOs$_2$O$_6$. 
Recently, it was reported that 
$T_1^{-1}$ of $^{39}$K nuclear spins in KOs$_2$O$_6$\cite{Takigawa}
exhibits an unusual temperature dependence in the normal state. 
This was interpreted as evidence for 
considerable antiferromagnetic correlations in the itinerant electron system. 
Assuming that both data sets are reliable, 
it must be concluded that the magnetic features of RbOs$_2$O$_6$ 
are quite different from those of KOs$_2$O$_6$. 

In the superconducting state, 
$T_1^{-1}$ reveals no clear coherence peak just below $T_c$ 
(= 3.8 K for $\mu_0H$ = 2.94 T) but drops sharply only below 3 K, 
i.e., at a temperature significantly lower than $T_c$. 
The maximum value of $(T_1T)^{-1}$ below $T_c$ is only 4 $\%$ 
larger than the value in the normal state. 
A similar result has recently been obtained by other workers\cite{Takigawa}. 
This behavior is distinctly different from that of 
well identified unconventional superconductors 
where $T_1^{-1}$ drops sharply just below $T_c$. 
We argue that the superconducting state of RbOs$_2$O$_6$ is 
of conventional type and that the data between 3 K and $T_c(H)$ reflects 
a strongly reduced coherence peak in $T_1^{-1}(T)$. 
The coherence peaks for $s$-wave type-II superconductors 
are often reduced for various reasons. 
In the case of V$_3$Sn\cite{MacLaughlin}, 
it was argued that the application of an external magnetic field 
causes the observed reduction. 
The coherence peak may also be suppressed by finite life-time effects 
on the quasiparticles due to, for instance, 
electron-phonon interactions\cite{Fibich}. 
Fibich showed that 
an effective broadening of the electronic energy levels is brought about by 
absorption processes of thermal phonons. 
More accurately, the gap function has a (negative) imaginary part, 
which has the effect of removing the singularity 
in the electrical density of states at non-zero temperature. 
Although these life-time effects are expected to be particularly important 
for strong-coupling superconductors, 
the mechanism was originally invoked in order to explain 
the only modest enhancement of $T_1^{-1}$ below $T_c$ in the case of Al. 
Fibich has derived the ratio $T_{1n}/T_{1s}$ of 
the relaxation rates in the superconducting state to those in the normal state 
as 
\begin{equation}
\frac{T_{1n}}{T_{1s}} = 2f(\Delta_1)[1+\frac{\Delta_1(T)}{k_{\rm B}T}(1-f(\Delta_1))\ln(\frac{2\Delta_1(T)}{|\Delta_2(T)|})] 
\label{eq:Fibich-1}
\end{equation}
with 
\begin{equation}
\frac{\Delta_2(T)}{\Delta_0} = C[\frac{\Delta_0}{\Delta_1(T)}]^{1/3}(\frac{T}{T_c})^{8/3}
\label{eq:Fibich-2}
\end{equation}
where $\Delta_1(T)$ and $\Delta_2(T)$ 
are the real and the imaginary part of gap function, 
$f$ is the Fermi distribution function, and $C$ is a fitting parameter. 
Our simulations (data not shown) using Eq. (\ref{eq:Fibich-1}) 
do not yield satisfactory results for the case of RbOs$_2$O$_6$, 
however, and we conclude that this type of reasoning 
is not adequate for explaining the only weakly developed coherence peak 
in $T_1^{-1}(T)$ in RbOs$_2$O$_6$. 

In an attempt to elucidate the reduction of the coherence peak in $T_1^{-1}$ 
below $T_c$, 
we tried to consider the effect of the applied magnetic field 
using the approach suggested by Goldberg and Weger\cite{Goldberg}. 
Here, the basic assumption is that 
the total nuclear spin-lattice relaxation rate is the sum of two terms, 
where the first describes the relaxation in the normal-state vortex cores 
and the other captures $T_1^{-1}$ in the remaining superconducting volume. 
Near $T_c$ this leads to 
\begin{equation}
(T_1T)^{-1} = (T_1T)_n^{-1}\frac{H\xi^2}{\Phi} + (T_1T)_{BCS}^{-1}(1-\frac{H\xi^2}{\Phi}) 
\label{eq:Goldberg}
\end{equation}
where $(T_1T)_n^{-1}$ = 0.117 (sK)$^{-1}$. 
The coherence length is given by 
$\xi(T) = \frac{0.74\xi(0)}{(1-\frac{T}{T_c(H)})^{1/2}}$ 
with $\xi(0) = 74 \AA$\cite{Rb-2}, and 
$(T_1T)_{BCS}^{-1}$ represents the relaxation in the superconducting volume. 
Inserting the parameters for RbOs$_2$O$_6$, 
the first term on the {\it r.h.s.} of Eq. (\ref{eq:Goldberg}) 
turns out to be negligibly small 
and hence for the field strength used in our experiments, 
the influence of the applied magnetic field 
in the manner described above plays no role in our problem. 
This may not be the case for higher applied magnetic fields, however. 

At very low temperatures and high magnetic fields, yet 
another process, namely spin diffusion, will play a dominant role. 
In this case, the observed relaxation rate $T_1^{-1}$ 
will be dominated by fast processes in the normal core of the vortices. 
Note that due to the small value of $\xi$ at low temperatures, 
finite-size effects may have to be taken into account in the vortex cores 
and the corresponding relaxation rate may not simply be given by $(T_1)_n^{-1}$ 
of the bulk in the normal state. 

Below 3 K, $T_1^{-1}$ decreases exponentially with temperature upon cooling. 
Below 1 K, where $T_1^{-1}$ is less than 7$\times$10$^{-3} s^{-1}$, 
the relaxation tends towards a temperature-independent value 
with decreasing temperature. 
This deviation may be caused by paramagnetic impurities 
and/or by relaxation via spin diffusion to the normal vortex cores, 
which are regions of fast relaxation. 
The onset of anomalous relaxation follows 
a more than an order of magnitude reduction of $T_1^{-1}$, 
indicating that the bulk of the sample ($\geq 95 \%$) 
is indeed superconducting. 
In our subsequent analysis, 
we refrain from discussing the data for $T \leq$ 1 K.

The inset of Fig. 3 shows a semilogarithmic plot of 
$T_1^{-1}$ vs $\frac{1}{T}$, which confirms the exponential decay of $T_1$ 
according to $T_1^{-1} \propto \exp(-\frac{\Delta(0)}{k_{\rm B}T})$ 
at temperatures well below $T_c$. 
The slope of the solid line implies that 
$\frac{\Delta(0)}{k_{\rm B}}$ = 6.1 K. 
With $T_c$ = 3.8 K, we obtain $\frac{2\Delta(0)}{k_{\rm B}T_c}$ = 3.2. 
This is in fair agreement with the value of the conventional BCS theory 
in the weak-coupling regime, $\frac{2\Delta(0)}{k_{\rm B}T_c}$ = 3.5. 
This finding, however, should be considered with some caution 
because, as we shall see below, 
a more detailed inspection of the data suggest 
a considerable anisotropy in the gap parameter. 
Nevertheless, the observed thermally activated temperature dependence of 
$T_1^{-1}$ provides clear evidence for a nodeless gap configuration 
and the remnants of the coherence peak just below $T_c$ 
indicate a conventional $s$-wave-type pairing of the quasiparticles. 
This is consistent with previous interpretations of results of 
specific-heat\cite{Rb-2} 
and magnetic-field penetration depth measurements\cite{Rb-3}. 

In order to reproduce $T_1^{-1}(T)$ at intermediate temperatures below $T_c$, 
we fitted the data to the BCS model 
and assumed a distribution of energy gap amplitudes 
in the range between $\Delta-\delta$ and $\Delta+\delta$ 
across the Fermi surface\cite{Hebel}. 
The solid line below $T_c$ in the mainframe of Fig. 3 
represents the result of a calculation using 
$\frac{2\Delta(0)}{k_{\rm B}T_c}$ = 3.5 and $\frac{\delta}{\Delta(T)}$ = 0.5. 
In this way, we obtain good agreement with the experimental results 
at temperatures between 1 K and 3 K. 
The calculation cannot reproduce the behavior just below $T_c$ quantitatively. 
The distribution of the energy gap can be accounted for 
by incorporating a $k$-space anisotropy 
of the conventional $s$-wave gap\cite{MacLaughlin}, 
\begin{equation}
\Delta(H,\Omega) = <\Delta(H)>[1+a(\Omega)]
\end{equation}
where $\Omega$ is the solid angle in $k$-space, 
$<\Delta(H)>$ the mean gap value over all orientations in $k$-space, 
and $a(\Omega)$ is the anisotropy function 
satisfying the condition $<a(\Omega)> = 0$. 
Our value for $\frac{\delta}{\Delta(T)}$ implies $<a^2(\Omega)>$ = 0.25. 
This anisotropy can substantially affect various thermodynamic quantities. 
For instance, it has been shown\cite{Clem-1,Clem-2} 
that it renormalizes the ratio $\frac{\Delta}{k_{\rm B}T_c}$ as 
$\frac{\Delta}{k_{\rm B}T_c} = \frac{\Delta^0}{k_{\rm B}T_c^0}(1-\frac{3}{2}<a^2>)$ where $\frac{\Delta^0}{k_{\rm B}T_c^0}$ is the ratio 
in the absence of anisotropy. In our case, 
since $\frac{2\Delta}{k_{\rm B}T_c}$ = 3.2 and $<a^2(\Omega)>$ = 0.25, 
it follows that $\frac{\Delta^0}{k_{\rm B}T_c^0}$ = 5.1, 
a substantially enhanced value 
with respect to the expectations for the BCS theory 
in the weak-coupling limit, 
and raising some questions about the validity of the approximation. 
An refined quantitative discussion of the NMR data obviously requires 
additional efforts in numerical calculations. 


%
\begin{figure}
\includegraphics[width=0.8\linewidth]{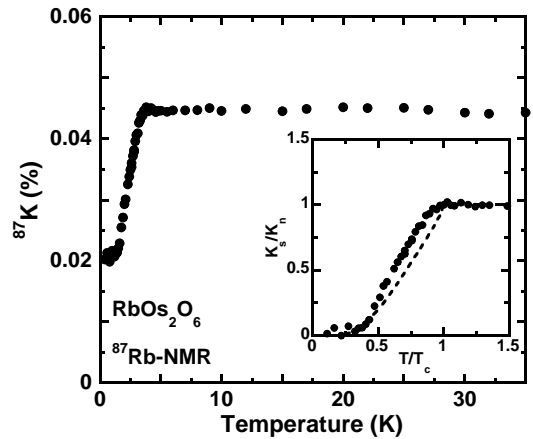}
\caption{\label{fig:epsart} 
Temperature dependence of the Knight shift of the $^{87}$Rb NMR signal.
Inset : Temperature dependence of the Knight shift $K_s$, 
normalized by its normal-state value at $T_c$. 
The dotted line is a calculation based on the conventional BCS theory 
with the same parameters 
as those employed in the analysis of $T_1^{-1}$ below $T_c$. 
}
\end{figure}

Figure 4 shows the temperature dependence of the Knight shift 
of the $^{87}$Rb NMR signal. 
The Knight shift is a measure of the uniform magnetic susceptibility 
of the conduction electrons, seen at a particular nuclear site. 
In general, the Knight shift consists of a $T$-independent orbital part 
and a spin part. 
In the normal state of RbOs$_2$O$_6$, 
the Knight shift is very small, about 0.045 $\%$, 
and practically independent of temperature, 
reflecting the influence of the temperature-independent contribution 
of the electronic spin susceptibility. 
In the superconducting state, 
the diamagnetic shift $H_{dia}$ is estimated as 
$-\frac{H_{c1}\ln(\frac{\beta d}{\sqrt{e}\xi})}{\ln(\frac{\lambda}{\xi})}$\cite{deGennes}. 
With $\xi = 74 \AA$\cite{Rb-2}, 
$\lambda = 4100 \AA$\cite{Rb-3}, 
$\beta$ = 0.381 for the triangular vortex lattice, and 
$d = 285 \AA$ as the nearest-neighbor vortex-lattice spacing in 2.9427 T, 
we calculate $K_{dia} = -0.004 \%$. 
The observed Knight shift variation below $T_c$ is much larger than $K_{dia}$, 
thus reflecting the spin-singlet-pairing of the quasiparticles. 
The residual line shift at very low temperatures 
is of the order of 0.020 $\%$. 

The inset of Fig. 4 shows the temperature dependence of 
the Knight shift $K_s$ normalized by its value at $T_c$. 
Here, the residual line shift attributed to orbital effects 
has been substracted from the raw data. 
The dotted line represents the result of a calculation 
using the conventional BCS model with the same parameters 
and the same gap distribution 
as those in the previously discussed analysis of $T_1^{-1}(T)$ below $T_c$. 
Because the calculation does not agree with the experimental data, 
the deviation of the temperature dependence of the Knight shift below $T_c$ 
from the BCS expectation needs further examination. 

The ratio $K_{\alpha} = \frac{S}{T_1TK_s^2}$ in the normal state 
provides a useful measure for the importance of 
electron-electron magnetic correlations\cite{Moriya,Narath}. 
The parameter $S = \frac{\gamma_e^2}{\gamma_n^2}\frac{\hbar}{4\pi k_{\rm B}}$ 
and $\gamma_e$ and $\gamma_n$ are 
the electronic and nuclear gyromagnetic ratios, respectively. 
Depending on the value of $K_{\alpha}$ 
being much smaller or larger than unity, 
substantial ferro- or antiferro-magnetic correlations 
in the itinerant electron systems are significant. 
If we assume that the residual line shift $K(T = 0)$ 
is due to the orbital contributions, 
$K_{\alpha}$ is estimated to be 4.6, 
providing some evidence for the existence of 
antiferromagnetic electron-electron correlations. 


We present and discuss the results of $^{87}$Rb NMR measurements 
on the new superconducting pyrochlore-type oxide RbOs$_2$O$_6$. 
In the normal state, 
the nuclear spin lattice relaxation rate $T_1^{-1}$ 
obeys the Korringa-type relation $(T_1T)_n^{-1}$ = 0.117 (sK)$^{-1}$ 
and the line shift is independent of temperature. 
In the superconducting state, 
$T_1^{-1}$ reveals a very much reduced coherence peak just below $T_c$, 
and eventually decreases with a thermally activated behavior 
upon further cooling. 
The $T_1^{-1}(T)$ data can qualitatively be explained by the BCS model 
considering some anisotropy of the gap function. 
In spite of some remaining numerical inconsistencies, 
we claim that our NMR results imply that 
the superconducting state of RbOs$_2$O$_6$ 
is characterized by singlet-pairing of the electrons 
and that the gap function exhibits a conventional $s$-wave-type symmetry; 
with some $k$-dependent variation of the amplitude, however. 

This study was partly supported by the Swiss National Science Foundation.

\end{document}